\documentclass[aps,prd,showpacs,onecolumn,preprintnumbers,notitlepage,nofootinbib,amsfont,amssymb,10pt,singlespacing] {revtex4-1}
\usepackage{amsmath,bm}
\usepackage{times}
\usepackage{braket}
\usepackage{color,graphicx}
\usepackage{slashed}
\usepackage{CJK,upgreek,fancyhdr}
\usepackage{hyperref}

\newcommand{\beq}{\begin{eqnarray}}
\newcommand{\eeq}{\end{eqnarray}}
\newcommand{\non}{\nonumber\\ }

\def\lsim{ {\ \lower-1.2pt\vbox{\hbox{\rlap{$<$}\lower6pt\vbox{\hbox{$\sim$}
}}}\ } }
\def\gsim{ {\ \lower-1.2pt\vbox{\hbox{\rlap{$>$}\lower6pt\vbox{\hbox{$\sim$}
}}}\ } }

\def \jhep{ J. High Energy Phys.  }

\definecolor{Red}{rgb}{1.,0.,0.}

\definecolor{Blue}{rgb}{0.,0.,1.}

\definecolor{nicered}{rgb}{0.7,0.1,0.2}
\definecolor{nicegreen}{rgb}{0.1,0.4,0.2}

\hypersetup{colorlinks,citecolor=nicegreen,linkcolor=nicered}

\begin{document}
%%%%%%%%%%%%%%%%%%%%%%%%%%%%%%%%%%%%%%%%%%%%%
%%
\title{\boldmath
Hadronic decays of $B \to a_1(1260) b_1(1235)$ in the perturbative QCD approach }
%===========================================
\author{Hao-Yang~Jing}
%\email{2020150648@jsnu.edu.cn}

\author{Xin~Liu\footnote{Corresponding author}
}
\email{liuxin@jsnu.edu.cn}

\affiliation{\small School of Physics and Electronic Engineering,
Jiangsu Normal University, Xuzhou 221116, China}

%%%%%%%%%%%%%%%%%%%%%%%%%%%%%%%%%%%%%%%%%%%%%%%%%%%%%%%%%%%%%%%%%%%%%

\author{Zhen-Jun~Xiao}
\email{xiaozhenjun@njnu.edu.cn}

\affiliation{\small Department of Physics and Institute of Theoretical
Physics, Nanjing Normal University, Nanjing 210023, China}
\date{\today}
%%============================================%%

\begin{abstract}
We calculate the branching ratios and polarization
fractions of the $B \to a_1 b_1$ decays in the perturbative QCD(pQCD)
approach at leading order, where $a_1$($b_1$) stands for the
axial-vector $a_1(1260)[b_1(1235)]$ state. By combining the
phenomenological analyses with the perturbative calculations,
we find the following results:
(a) the large decay rates around $10^{-5}$ to $10^{-6}$ of the
$B \to a_1 b_1$ decays dominated by the longitudinal
polarization(except for the $B^+ \to b_1^+ a_1^0$ mode) are predicted
and basically consistent with those
in the QCD factorization(QCDF) within errors,
which are expected to be tested by
the Large Hadron Collider
and Belle-II experiments. The large $B^0 \to a_1^0 b_1^0$ branching
ratio could provide hints to help explore the mechanism of
the color-suppressed decays.
(b) the rather different QCD behaviors between the
$a_1$ and $b_1$ mesons result in the destructive(constructive)
contributions in the nonfactorizable spectator diagrams with
$a_1(b_1)$ emission. Therefore, an interesting pattern of the
branching ratios appears for the color-suppressed
$B^0 \to a_1^0 a_1^0, a_1^0 b_1^0,$ and
$b_1^0 b_1^0$ modes in the pQCD approach,
$Br(B^0 \to b_1^0 b_1^0) >
Br(B^0 \to a_1^0 b_1^0) \gtrsim Br(B^0 \to a_1^0 a_1^0)$,
which is different from
$Br(B^0 \to b_1^0 b_1^0)
\sim Br(B^0 \to a_1^0 b_1^0) \gtrsim Br(B^0 \to a_1^0 a_1^0)$
in the QCDF and would
be verified
at future experiments.
(c) the large naive factorization breaking effects are
observed in these $B \to a_1 b_1$ decays.
Specifically, the large nonfactorizable spectator(weak annihilation)
amplitudes contribute to the $B^0 \to b_1^+ a_1^-(B^+ \to a_1^+ b_1^0\;
{\rm and}\; B^+ \to b_1^+ a_1^0)$ mode(s), which demand confirmations
via the precise measurements.
Furthermore, the different phenomenologies shown among $B \to a_1 b_1$,
$B \to a_1 a_1$, and $B \to b_1 b_1$ decays are also expected to be tested
stringently,
which could shed light on the typical QCD
dynamics involved in these modes,
even further distinguish those two
popular pQCD and QCDF approaches.

\end{abstract}

\pacs{13.25.Hw, 12.38.Bx, 14.40.Nd}
\preprint{\footnotesize JSNU-PHY-HEP-2017-2}
\maketitle

%\section{Introduction}

It is well known that the nonleptonic $B$ meson decays can provide
highly important information to understand the physics within and/or
beyond the standard model(SM). Specifically, they can help
us to study the perturbative and non-perturbative quantum chromodynamics(QCD),
search for the charge-parity(CP) violation to further find out its origin,
determine the Cabibbo-Kobayashi-Maskawa(CKM) phases $\alpha(\phi_2), \beta(\phi_1)$, and $\gamma(\phi_3)$ in the unitary triangle,
even identify the possible new physics hidden in the higher energy scale, etc.
Moreover, one can also indirectly conjecture the inner
structure of the hadrons involved in the final states through the precise
measurements experimentally. The great efforts have been extensively contributed
to the exclusive $B \to PP, PV,$ and $VV$ decays at both theoretical and experimental
aspects in the past decades, for example, see Refs.~\cite{Wirbel:1985ji,Bauer:1986bm,Ali:1997nh,Ali:1998eb,Du:2001hr,
Beneke:2003zv,Beneke:2006hg,Li:2006jv,Ali:2007ff,Wang:2008rk,
Cheng:2009cn,Zou:2015iwa,Zhou:2015jba,Zhou:2016jkv,Wang:2017hxe,
Wang:2017rmh,Olive:2016xmw,Amhis:2016xyh}, where $P$ and $V$ denote
the $S$-wave pseudoscalar
and vector states, respectively. However, the known "puzzles", for example,
the large observed
$B^0 \to \pi^0 \pi^0$, $B^0 \to \rho^0 \pi^0$, and $B \to K \eta^\prime$
decay rates,
the experimental inequality of the direct {\it CP} asymmetries
between $B^\pm \to K^\pm \pi^0$ and $B^0 \to K^\pm \pi^\mp$ modes, the unknown
mechanism of the polarization in the penguin-dominated $B \to VV$ processes etc.,
are still not elegantly resolved~\cite{Cheng:2009xz,Olive:2016xmw,Amhis:2016xyh}.
Therefore, a large variety of
relevant $B$ meson decay modes
should be opened to help us get deep understanding complementarily.

Fortunately, two successful $B$-factory experiments, i.e.,
{\it BABAR} at SLAC and Belle at KEK, have measured many
nonleptonic $B$ meson decays into the final states containing $p$-wave light hadrons in the last decade~\cite{Olive:2016xmw,Amhis:2016xyh}. Then the Large Hadron Collider-beauty(LHCb) experiments
at CERN almost became the only apparatus to explore the physics of $b$ quark in recent years. A large number of data related to nonleptonic $B$ decays have been reported~\cite{Olive:2016xmw,Amhis:2016xyh}.
The forthcoming start of the upgraded Belle-II experiment will further
improve the measurements. The Future Circular Collider and Circular Electron-Positron Collider are expected to give further chance for the studies on $B$ meson decays~\cite{CEPC}. Therefore, it is believed that the great supports coming from these current running
and forthcoming experiments could dramatically promote our understanding of the nature.

In this work, we will study the nonleptonic charmless decays of $B \to a_1(1260) b_1(1235)$ in the SM. For the sake of simplicity, the abbreviation $a_1$ and $b_1$ will be used in the following content to denote the $a_1(1260)$ and $b_1(1235)$ mesons, respectively, unless otherwise stated.
As we know, the considered processes contain the same
components as the $B \to \pi \pi, \rho \pi, \rho\rho$ modes at the quark level.
The latter decays have contributed to the determination and constraints on the CKM angle $\alpha$~\cite{Olive:2016xmw}.
Certainly, the $B \to a_1(b_1) \pi, a_1(b_1) \rho$, and $a_1(b_1) b_1(a_1)$
decays can also
provide useful information to the angle $\alpha$ complementarily~\cite{Lombardo:2009kt,Aubert:2009ab,
Cheng:2007mx,Wang:2008hu,Cheng:2008gxa,Zhang:2012ew,Liu:2012jb}. Particularly,
because $a_1$ and $b_1$ behave
differently from each other, these considered decays
could provide opportunities for us
to explore the interesting QCD dynamics. Furthermore, the $B \to a_1 b_1$ decays
with $b_1$ emission could provide more evidence for probing the naive factorization
breaking effects~\cite{Diehl:2001xe} because the decay constant $f_{b_1}$ vanishes owing to the charge conjugation
invariance for the neutral $b_1^0$ state or the even G-parity validity
in the isospin limit for the charged $b_1^\pm$ states~\cite{Yang:2005gk,Yang:2007zt,Cheng:2007mx}.

As stated in the naive factorization hypothesis~\cite{Bauer:1986bm},
the hadronic matrix element of a $B$ meson decay amplitude can be
expressed by the factorizable emission amplitudes as
a production of the decay constants and the transition
form factors. Then, for example, the $B^0 \to b_1^+ a_1^-$
mode with $b_1$ emission almost receives no factorizable
contributions due to the vanishing decay constant $f_{b_1}$ and the
branching ratio would approach
to zero in the naive factorization. While, it is worth emphasizing
that the corresponding decay rate predicted in the QCD
factorization(QCDF)~\cite{Beneke:1999br,Du:2001hr} by
including the nonfactorizable spectator and
annihilation contributions can reach ${\cal O}(10^{-6})$
\cite{Cheng:2008gxa}, which is detectable at the
current experiments. It means that these important contributions violate the
naive factorization
if this large decay rate would be confirmed by the related experiments.
However, because of the unavoidable endpoint singularities, the nonfactorizable
spectator amplitudes, as well as the annihilation
ones, have to be determined by data fitting accompanied with large uncertainties
in the framework of QCDF~\cite{Beneke:1999br,Beneke:2003zv}. Luckily, the perturbative QCD(pQCD) approach~\cite{Keum:2000ph,Lu:2000em}, which bases on the framework of $k_T$ factorization theorem, is appropriate to calculate the decay amplitudes with the nonfactorizable spectator and annihilation topologies. Since it keeps the transverse momentum $k_T$ of the valence quark in the hadrons, then the resultant Sudakov
factor[$e^{(-S)}$] and threshold factor[$S_t(x)$], which smear the endpoint singularities, make the pQCD approach more self-consistent.
More details about this pQCD approach can be found in the review paper~\cite{Li:2003yj}.

We will therefore study the
branching ratios and polarization fractions of the considered
$B \to a_1 b_1$ decays in the pQCD approach, with which the nonfactorizable spectator and
annihilation Feynman diagrams can be calculated perturbatively. It is worth stressing that
the observations of the pure annihilation $B_s^0 \to \pi^+ \pi^-$ and $B_d^0 \to K^+ K^-$ decays performed by the CDF~\cite{Ruffini:2013jea}
and LHCb~\cite{Aaij:2012as} collaborations have confirmed the pQCD calculations~\cite{Li:2004ep,Ali:2007ff,Xiao:2011tx} of the annihilation type diagrams~\footnote{Certainly, the soft-collinear
effective theory(SCET)~\cite{Bauer:2000yr} has a
different point of view on the calculations of the annihilation
diagrams~\cite{Arnesen:2006vb} . We believe that
this discrepancy between the pQCD
approach and SCET
could be finally
resolved through the precise measurements experimentally. Therefore,
this conversation will be put aside in the present work.}. Moreover,
both of $a_1$ and $b_1$ are axial-vector($A$)
states but with different quantum numbers $J^{PC}=1^{+-}$ and $1^{++}$ correspondingly.
It is believed that the $B \to AA$ decays could
provide more information on the helicity structure of
the decay mechanism because, like $B \to VV$ decays, they
also contain three polarization states~\cite{Cheng:2008gxa}, which would be helpful to
understand the famous "polarization puzzle" in a different way.

%\section{Theoretical Framework} \label{sec:1}

For the considered $B \to a_1 b_1$ decays with $\bar{b} \to \bar{d}$ transition,
the related weak effective
Hamiltonian $H_{eff}$~\cite{Buchalla:1995vs} can be written as
\begin{equation}
H_{eff}\, =\, {G_F\over\sqrt{2}}
\left\{V_{ub}^*V_{ud} \left[C_1(\mu)O_1^{u}(\mu)
+C_2(\mu)O_2^{u}(\mu)\right]- V_{tb}^*V_{td} \sum_{i=3}^{10}C_i(\mu)O_i(\mu)\right\}\;,
\label{eq:heff}
\end{equation}
with the Fermi constant $G_F=1.16639\times 10^{-5}{\rm
GeV}^{-2}$, CKM matrix elements $V$,
and Wilson coefficients $C_i(\mu)$ at the renormalization scale
$\mu$. The local four-quark
operators $O_i(i=1,\cdots,10)$ are written as
\begin{enumerate}
\item[]{(1) current-current (tree) operators}
\beq
\begin{split}
O_1^{u}\, =\,(\bar{d}_\alpha u_\beta)_{V-A}(\bar{u}_\beta b_\alpha)_{V-A}\;,
O_2^{u}\, =\, (\bar{d}_\alpha u_\alpha)_{V-A}(\bar{u}_\beta b_\beta)_{V-A}\;;
\end{split}
\eeq
\item[]{(2) QCD penguin operators}
\beq
\begin{split}
O_3\, =\, (\bar{d}_\alpha b_\alpha)_{V-A}\sum_{q'}(\bar{q}'_\beta q'_\beta)_{V-A}\;,
O_4\, =\, (\bar{d}_\alpha b_\beta)_{V-A}\sum_{q'}(\bar{q}'_\beta q'_\alpha)_{V-A}\;,
 \\
O_5\, =\, (\bar{d}_\alpha b_\alpha)_{V-A}\sum_{q'}(\bar{q}'_\beta q'_\beta)_{V+A}\;,
O_6\, =\, (\bar{d}_\alpha b_\beta)_{V-A}\sum_{q'}(\bar{q}'_\beta q'_\alpha)_{V+A}\;;
\end{split}
\eeq
\item[]{(3) electroweak penguin operators}
\beq
\begin{split}
O_7\, =\,
\frac{3}{2}(\bar{d}_\alpha b_\alpha)_{V-A}\sum_{q'}e_{q'}(\bar{q}'_\beta q'_\beta)_{V+A}\;,
O_8\, =\,
\frac{3}{2}(\bar{d}_\alpha b_\beta)_{V-A}\sum_{q'}e_{q'}(\bar{q}'_\beta q'_\alpha)_{V+A}\;,
\\
O_9\, =\,
\frac{3}{2}(\bar{d}_\alpha b_\alpha)_{V-A}\sum_{q'}e_{q'}(\bar{q}'_\beta q'_\beta)_{V-A}\;,
O_{10}\, =\,
\frac{3}{2}(\bar{d}_\alpha b_\beta)_{V-A}\sum_{q'}e_{q'}(\bar{q}'_\beta q'_\alpha)_{V-A}\;.
\end{split}
\eeq
\end{enumerate}
with the color indices $\alpha, \; \beta$(not to be confused with the CKM angles)
and the notations
$(\bar{q}'q')_{V\pm A} = \bar q' \gamma_\mu (1\pm \gamma_5)q'$.
The index $q'$ in the summation of the above operators runs
through $u,\;d,\;s$, $c$, and $b$.
We will use the leading order Wilson coefficients to keep the consistency
since the calculations in this work are at leading order[${\cal O}(\alpha_s)$]
of the pQCD approach.
For the renormalization group evolution of the Wilson coefficients
from higher scale to lower scale, we use the formulas as given in
Ref.~\cite{Keum:2000ph} directly.

%%=========================================================================
\begin{figure}[!!b]
\centering
\begin{tabular}{l}
\includegraphics[width=0.8\textwidth]{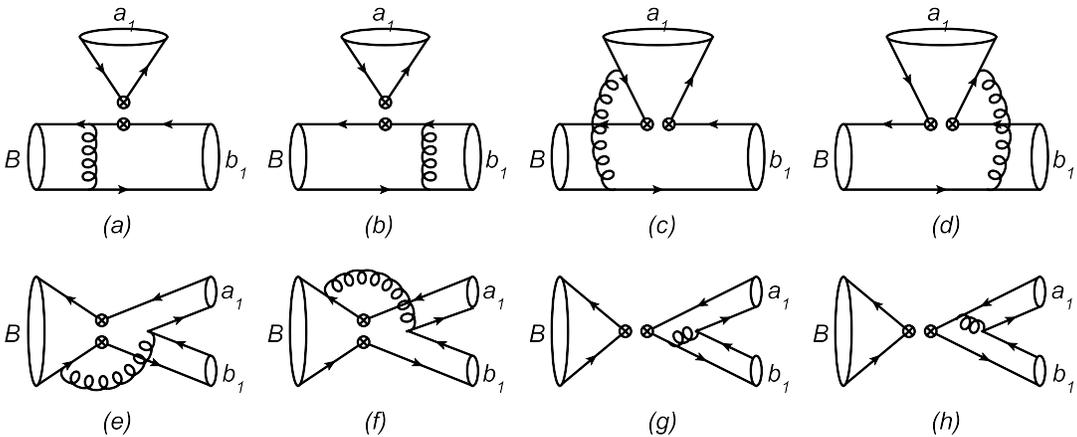}
\end{tabular}
\caption{ Typical Feynman diagrams for $B \to a_1 b_1$ decays at leading
order in the pQCD approach. By exchanging the position of the $a_1$
and $b_1$ mesons, one will obtain another eight Feynman diagrams that
possibly contribute to the considered $B \to a_1 b_1$ modes. }
\label{fig:fig1}
\end{figure}
%%%=========================================================================

%\section{Perturbative Calculations in pQCD approach} \label{sec:2}

Similar to the vector meson, the axial-vector one also has three kinds of polarizations,
i.e., longitudinal ($L$), normal ($N$), and transverse ($T$), respectively.
Therefore, analogous to the $B \to VV$
decays,
the $B\to a_1 b_1$ decay amplitudes will be characterized by the polarization states
of these axial-vector mesons.
In terms of helicities, the decay amplitudes ${\cal M}^{(\sigma)}$ for $B
\to a_1(P_2,\epsilon^*_2) b_1(P_3,\epsilon^*_3)$ decays can be
generally described by
\beq
{\cal M}^{(\sigma)}&=&\epsilon_{2\mu}^{*}(\sigma)\epsilon_{3\nu}^{*}(\sigma)
\left[ a \,\, g^{\mu\nu} + {b \over m_{a_1} m_{b_1}} P_1^\mu P_1^\nu + i{c
\over m_{a_1} m_{b_1} } \epsilon^{\mu\nu\alpha\beta} P_{2\alpha}
P_{3\beta}\right]\;,\non &\equiv &m_{B}^{2}{\cal
M}_{L}+m_{B}^{2}{\cal M}_{N}
\epsilon^{*}_{2}(\sigma=T)\cdot\epsilon^{*}_{3}(\sigma=T) \non &&
+i{\cal M}_{T}\epsilon^{\alpha \beta\gamma \rho}
\epsilon^{*}_{2\alpha}(\sigma)\epsilon^{*}_{3\beta}(\sigma)
P_{2\gamma }P_{3\rho }\; , \label{eq:msigma}
\eeq
where the superscript $\sigma$ denotes the helicity states of two mesons with $L(T)$
standing for the longitudinal (transverse) component and the
definitions of the amplitudes ${\cal M}_{h} (h=L,N,T)$ in terms
of the Lorentz-invariant amplitudes $a$, $b$ and $c$ are
\beq
m_{B}^2 \,\, {\cal M}_L &=& a \,\, \epsilon_2^{*}(L) \cdot
\epsilon_3^{*}(L) +{b \over m_{a_1} m_{b_1} } \epsilon_{2}^{*}(L) \cdot
P_3 \,\, \epsilon_{3}^{*}(L) \cdot P_2\;, \non
m_{B}^2 \,\, {\cal M}_N &=& a \;,\label{eq:amp}\\
m_{B}^2 \,\, {\cal M}_T &=& {c \over r_2\,
r_3}\;.\label{id-rel}\nonumber
\eeq
with $\epsilon_{2(3)}$ and $P_{2(3)}$ denoting the polarization vector
and momentum of the $a_1(b_1)$ state correspondingly. Here, $r_{2(3)}\equiv m_{a_1(b_1)}/m_B$
with $m_{a_1(b_1)}$ and $m_B$, the masses of the light $a_1(b_1)$ and heavy $B$ mesons, respectively.
We will therefore analyze the
helicity amplitudes ${\cal M}_L, {\cal M}_N, {\cal M}_T$ based on
the pQCD approach.
According to the helicity amplitudes~(\ref{eq:amp}), the
transversity ones can be defined as
\beq
{\cal A}_{L}&=&\xi
m^{2}_{B}{\cal M}_{L}, \quad {\cal A}_{\parallel}=\xi
\sqrt{2}m^{2}_{B}{\cal M}_{N}, \quad {\cal A}_{\perp}=\xi r_2 r_3
\sqrt{2(r^{2}-1)} m^{2}_{B} {\cal M }_{T}\;. \label{eq:ase}
\eeq
for the longitudinal, parallel, and perpendicular polarizations,
respectively, where the
ratio $r=P_{2}\cdot P_{3}/(m_{B}^2\; r_2 r_3)$ and the normalization factor
$\xi=\sqrt{G^2_{F}{\bf{P_c}} /(16\pi m^2_{B}\Gamma)}$ with the decay width
$\Gamma =\frac{G_{F}^{2}|\bf{P_c}|}{16 \pi m^{2}_{B} }
\sum_{\sigma}{\cal M}^{(\sigma)\dagger }{\cal M^{(\sigma)}}$ and the momentum of either of the
outgoing axial-vector mesons $|\bf{P_c}|\equiv |\bf{P_{2z}}|=|\bf{P_{3z}}|$. These amplitudes satisfy
the following relation,
\beq |{\cal A}_{L}|^2+|{\cal
A}_{\parallel}|^2+|{\cal A}_{\perp}|^2=1\;.
\label{eq:norm}
\eeq

As illustrated in Fig.~\ref{fig:fig1}, analogous to the $B \to a_1 a_1$ and $b_1 b_1$ decays~\cite{Liu:2012jb},
there are 8 types of diagrams contributing to the $B \to a_1 b_1$ decays
at the lowest order in the pQCD approach.
Because the amplitudes for the Feynman diagrams of the $B \to AA$ decays have been
analyzed explicitly in Ref.~\cite{Liu:2012jb}, then the $B \to a_1 b_1$ decay amplitudes
can be easily obtained from the Eqs.~(25)-(60) by appropriate replacements correspondingly:
\begin{itemize}
\item[]{(1)}
When the $a_1(b_1)$ state flies(recoils) along with the $+z(-z)$ direction in the
$B$ meson rest frame, the above mentioned Eqs.~(25)-(60)~\cite{Liu:2012jb}
will describe the
$B \to a_1 b_1$ decays with $B \to b_1$ transition, in which the related
$B \to b_1$ form factor can be factored out. The Feynman decay amplitudes
will be
expressed with $F^{h}$ and $M^{h}$;

\item[]{(2)}
When the $b_1(a_1)$ state flies(recoils) along with the $+z(-z)$ direction in the
$B$ meson rest frame, the above mentioned Eqs.~(25)-(60)~\cite{Liu:2012jb} will describe the
$B \to a_1 b_1$ decays with $B \to a_1$ transition, in which the related
$B \to a_1$ form factor can also be extracted out. The Feynman decay
amplitudes will be presented with $F^{\prime h}$ and $M^{\prime h}$.
\end{itemize}
Hence, for simplicity, we will not present the factorization formulas for these
$B \to a_1 b_1$
modes again in this work. The interested readers can refer to Ref.~\cite{Liu:2012jb} for details. By combining various contributions from the relevant Feynman diagrams together,
the decay amplitudes of the $B \to a_1 b_1$ decays can then be collected straightforwardly with three polarizations $h=L, N, T$ as follows:
\beq
{\cal M}_h(B^0 \to a_1^+ b_1^-) &=&
\xi_u \biggl[ a_1 F_{fs}^h + C_1 M_{nfs}^h + C_2 M_{nfa}^{\prime h}
+ a_2 f_{B} F_{fa}^{\prime h} \biggr] - \xi_t \biggl[ (a_4+ a_{10}) F_{fs}^h
+ (C_3 + C_9) M_{nfs}^h
\non &&
+ (C_5+ C_7) M_{nfs}^{h,P_1} + (C_3 + C_4 -\frac{1}{2}
(C_9 + C_{10})) M_{nfa}^{h} + (C_4+ C_{10}) M_{nfa}^{\prime h}
\non  & &
+ (C_5 - \frac{1}{2} C_7) M_{nfa}^{h,P_1} + ( C_6 - \frac{1}{2} C_8) M_{nfa}^{h,P_2}
+ ( a_3 + a_4  + a_5 - \frac{1}{2} (a_7 + a_9 + a_{10}) ) f_B F_{fa}^{h}
\non &&
+ ( C_6 + C_8) M_{nfa}^{\prime h,P_2}
+ ( a_3 + a_5  + a_7 + a_9 )f_B F_{fa}^{\prime h} + (a_6
- \frac{1}{2} a_8) f_B F_{fa}^{h,P_2}   \biggr]\;, \label{eq:tda-b02a1pb1m1}
\eeq
\beq
{\cal M}_h(B^0 \to b_1^+ a_1^-) &=&
\xi_u \biggl[ a_1 F_{fs}^{\prime h} + C_1 M_{nfs}^{\prime h} + C_2 M_{nfa}^h
+ a_2 f_{B} F_{fa}^h \biggr] - \xi_t \biggl[ (a_4+ a_{10}) F_{fs}^{\prime h}
+ (C_3 + C_9) M_{nfs}^{\prime h}
\non &&
+ (C_5+ C_7) M_{nfs}^{\prime h,P_1} + (C_3 + C_4 -\frac{1}{2}
(C_9 + C_{10})) M_{nfa}^{\prime h} + (C_4+ C_{10}) M_{nfa}^h
\non  & &
+ (C_5 - \frac{1}{2} C_7) M_{nfa}^{\prime h,P_1} + ( C_6 - \frac{1}{2} C_8) M_{nfa}^{\prime h,P_2}
+ ( a_3 + a_4  + a_5 - \frac{1}{2} (a_7 + a_9 + a_{10}) )f_B F_{fa}^{\prime h}
\non &&
+ ( C_6 + C_8) M_{nfa}^{h,P_2}
+ ( a_3 + a_5  + a_7 + a_9 )f_B F_{fa}^{h} + (a_6
- \frac{1}{2} a_8) f_B F_{fa}^{h,P_2}   \biggr]\;, \label{eq:tda-b02b1pa1m}
\eeq
\beq
\sqrt{2} {\cal M}_{h}(B^+ \to a_1^+ b_1^0) &=&
\xi_u \biggl[a_1( F_{fs}^{h}- f_B F_{fa}^{\prime h} + f_B F_{fa}^{h})
+a_2 F_{fs}^{\prime h} + C_1(M_{nfs}^{h}
+ M_{nfa}^{h} - M_{nfa}^{\prime h})
-C_2 M_{nfs}^{\prime h} \biggr]
\non &&
- \xi_t \biggl[ (\frac{5}{3} C_9 + C_{10} -\frac{1}{2} a_8 - a_4)F_{fs}^{\prime h} +
(a_4 + a_{10})F_{fs}^{h}
+(\frac{1}{2}a_9 - C_3)M_{nfs}^{\prime h}
\non &&
+ (\frac{1}{2}C_7 - C_5)M_{nfs}^{\prime h,P_1}
+\frac{3}{2} C_8 M_{nfs}^{\prime h,P_2} + (C_3 + C_9)M_{nfs}^{h} + (C_5 + C_7)M_{nfs}^{h,P_1}
\non &&
+ (C_3 + C_9)(M_{nfa}^{h,P_1} - M_{nfa}^{\prime h,P_1})
+ (a_4 + a_{10})(f_B F_{fa}^{h} - f_B F_{fa}^{\prime h})
\non &&
+  (a_6 + a_8)(f_B F_{fa}^{h,P_2} - f_B F_{fa}^{\prime h,P_2})\biggr]\;, \label{eq:tda-bp2a1pb10}
\eeq
\beq
\sqrt{2} {\cal M}_{h}(B^+ \to b_1^+ a_1^0) &=&
\xi_u \biggl[a_1( F_{fs}^{\prime h}- f_B F_{fa}^{h}+ f_B F_{fa}^{\prime h})
+a_2 F_{fs}^{h}  + C_1(M_{nfs}^{\prime h} + M_{nfa}^{\prime h} - M_{nfa}^{h})
-C_2 M_{nfs}^{h} \biggr]
\non &&
- \xi_t \biggl[ (\frac{5}{3} C_9 + C_{10} -\frac{1}{2} a_8 - a_4)F_{fs}^{h} +
(a_4 + a_{10})F_{fs}^{\prime h}+(\frac{1}{2}a_9 - C_3)M_{nfs}^{h}
\non &&
+ (\frac{1}{2}C_7 - C_5)M_{nfs}^{h,P_1}
+\frac{3}{2} C_8 M_{nfs}^{h,P_2} + (C_3 + C_9)M_{nfs}^{\prime h} + (C_5 + C_7)M_{nfs}^{\prime h,P_1}
\non &&
+ (C_3 + C_9)(M_{nfa}^{\prime h,P_1} - M_{nfa}^{h,P_1})
+ (a_4 + a_{10})(f_B F_{fa}^{\prime h} - f_B F_{fa}^{h})
\non &&
+  (a_6 + a_8)(f_B F_{fa}^{\prime h,P_2} - f_B F_{fa}^{h,P_2})\biggr]\;, \label{eq:tda-bp2b1pa10}
\eeq
%%%
\beq
2 {\cal M}_{h}(B^0 \to a_1^0 b_1^0) &=& \xi_u \biggl[-a_2
(F_{fs}^{\prime h}+ F_{fs}^{h} - f_B F_{fa}^{\prime h} - f_B F_{fa}^{h}) - C_2 (M_{nfa}^{\prime h} +
M_{nfa}^{h}- M_{nfs}^{\prime h} - M_{nfs}^{h})\biggr]
\non &&
-\xi_t \biggl[
(a_4 - \frac{1}{2} (3 a_7 + 3 a_9 + a_{10}) )(F_{fs}^{\prime h} + F_{fs}^{h})
- (C_5 - \frac{1}{2} C_7)
 (M_{nfs}^{\prime h,P_1} + M_{nfs}^{h,P_1})
 \non &&
+ (C_3 - \frac{1}{2}
(C_9 + 3 C_{10}))(M_{nfs}^{\prime h} + M_{nfs}^{h})+ (C_3+ 2 C_4 - \frac{1}{2} (C_9 - C_{10}))
(M_{nfa}^{\prime h} + M_{nfa}^{h})
\non &&  -\frac{3}{2}
C_8(M_{nfs}^{\prime h,P_2} + M_{nfs}^{h,P_2})
+ (2 a_3 + a_4 + 2 a_5 + \frac{1}{2} (a_7 - a_9 + a_{10}))(f_B F_{fa}^{\prime h} + f_B F_{fa}^{h}) \non &&  +
(C_5 - \frac{1}{2} C_7)(M_{nfa}^{\prime h,P_1} + M_{nfa}^{h,P_1})+ (2 C_6 + \frac{1}{2} C_8)
(M_{nfa}^{\prime h,P_2} + M_{nfa}^{h,P_2})
\non &&
+ (a_6 - \frac{1}{2} a_8)(f_B F_{fa}^{\prime h,P_2} + f_B F_{fa}^{h,P_2})\biggr]\;. \label{eq:tda-b02a10b10}
\eeq
where $\xi_u$ and $\xi_t$ stand for the products of CKM matrix elements $V_{ub}^* V_{ud}$ and $V_{tb}^* V_{td}$, respectively.
The standard combinations $a_i$ of Wilson coefficients are defined as follows,
  \beq
a_1&=& C_2 + \frac{C_1}{3}\;, \qquad  a_2 = C_1 + \frac{C_2}{3}\;,
\qquad  a_i = C_i + \frac{C_{i \pm 1}}{3}(i=3 - 10) \;.
  \eeq
where $C_2 \sim {\cal O}(1)$ and the upper(lower) sign applies when $i$ is odd(even).

%\section{Numerical Results and Discussions}\label{sec:3}

Now, we will turn to the numerical evaluations of the branching ratios and polarization
fractions of the considered $B \to a_1 b_1$ decays in the pQCD approach. The essential
comments on the input parameters are given as follows:
\begin{enumerate}

\item
For the heavy $B$ emsons and light axial-vector $a_1$ and $b_1$ states, the same hadron wave functions and distribution amplitudes including Gegenbauer moments are adopted as those in Ref.~\cite{Liu:2012jb}. And the same QCD scale, masses of hadrons, and decay constants are also utilized. The $B^0$ meson lifetime is updated as $1.52$~ps~\cite{Olive:2016xmw}.

\item
As for the CKM matrix elements, we adopt the Wolfenstein
parametrization at leading order and the newly updated parameters $A=0.811$,
 $\lambda=0.22506$, $\bar{\rho}=0.124$, and $\bar{\eta}=0.356$~\cite{Olive:2016xmw}.

\end{enumerate}

%%%========================================================================================================
%% ==========================================================================================================

%\subsection{CP-averaged Branching Ratios }\label{sec:3-a}

The theoretical predictions for $B \to a_1 b_1$ decays
evaluated in the pQCD approach,
together with the results in the QCDF approach,
have been grouped in the Tables~\ref{tab:Br-Pf-p0-00}-\ref{tab:Br-Pf-pmmp},
in which the first error is induced by the uncertainties of the shape parameter
$\omega_B=0.40 \pm 0.04$ GeV in the $B$ meson
wave function,
the second error arises from the combination of the uncertainties of Gegenbauer
moments $a_{2,a_1}^{\parallel}, a_{1,b_1}^{||}, a_{1,a_1}^{\perp}$, and $a_{2,b_1}^{\perp}$
in the distribution amplitudes of $a_1$ and $b_1$ mesons,
and the last error is also the combined uncertainty from the CKM matrix
elements: $\bar{\rho}=0.124^{+0.019}_{-0.018} $ and
$\bar{\eta}=0.356^{+0.011}_{-0.011}$ \cite{Olive:2016xmw}. It is easily seen
that the theoretical predictions suffer from large uncertainties that mainly
induced by the parameters describing the nonperturbative hadron dynamics.
It is therefore expected that the predictions given in the pQCD approach
could be improved greatly with the well-constrained inputs
based on the  nonperturbative QCD, e.g., Lattice QCD,  calculations
with high precision and/or the future precise measurements experimentally.

%%%========================================================================================================
\begin{table}[htb]
\caption{Branching ratios and polarization fractions of the $B^+ \to a_1^+ b_1^0, B^+ \to b_1^+ a_1^0$, and $B^0 \to a_1^0 b_1^0$ decays in the pQCD approach(This work).
For comparison, we also quote the related results predicted in the
QCDF approach~\cite{Cheng:2008gxa}.}
\label{tab:Br-Pf-p0-00}
 \begin{center}\vspace{-0.3cm}{\footnotesize
\begin{tabular}[t]{cc||cc|cc|cc}
\hline  \hline
   \multicolumn{2}{c||}{Decay Channels}   &  \multicolumn{2}{c|}{$B^+ \to a_1^+ b_1^0$}& \multicolumn{2}{c|}{$B^+ \to b_1^+ a_1^0$}&  \multicolumn{2}{c}{$B^0 \to a_1^0 b_1^0$} \\
 Parameter  & Definition & This work &   QCDF  & This work &   QCDF & This work &   QCDF\\
\hline
  BR($10^{-6}$)        & $\Gamma/ \Gamma_{\rm total}$
  &$9.0^{+1.5+5.2+0.8}_{-1.3-3.8-0.7}$&
  $37.8^{+23.9+11.4}_{-15.3-5.3}$&
  $4.2^{+0.4+2.0+0.4}_{-0.3-1.4-0.3}$&
  $1.0^{+1.6+6.2}_{-0.5-0.1}$&
  $3.3^{+0.6+1.8+0.3}_{-0.5-1.5-0.3}$&
  $3.8^{+6.2+2.6}_{-2.3-0.5}$\\
 $f_L$      & $|{\cal A}_L|^2$
 &$0.62^{+0.01+0.01+0.00}_{-0.03-0.04-0.00}$&
  $0.92^{+0.02}_{-0.24}$&
  $0.28^{+0.00+0.02+0.00}_{-0.01-0.04-0.01}$&
  $0.73^{+0.12}_{-0.82}$&
  $0.63^{+0.00+0.06+0.00}_{-0.01-0.10-0.01}$&
  $0.98^{+0.01}_{-0.31}$\\
 $f_{||}$   & $|{\cal A}_{||}|^2$
 &$0.10^{+0.00+0.05+0.00}_{-0.00-0.02-0.00}$&
 &$0.16^{+0.01+0.03+0.00}_{-0.02-0.04-0.02}$&
  &$0.17^{+0.00+0.04+0.01}_{-0.00-0.00-0.00}$&
  \\
 $f_{\perp}$& $|{\cal A}_\perp|^2$
 &$0.28^{+0.03+0.06+0.00}_{-0.02-0.03-0.00}$&
 &$0.57^{+0.02+0.04+0.01}_{-0.02-0.07-0.00}$& &
 $0.20^{+0.00+0.09+0.00}_{-0.00-0.07-0.00}$&
 \\ \hline \hline
\end{tabular}}
\end{center}
\end{table}
%% ==========================================================================================================

%%%========================================================================================================
\begin{table}[htb]
\caption{Same as Table~\ref{tab:Br-Pf-p0-00} but for the $B^0 \to a_1^+ b_1^-$
and $B^0 \to b_1^+ a_1^-$ decays. }
\label{tab:Br-Pf-pm-mp}
 \begin{center}\vspace{-0.5cm}{\footnotesize
\begin{tabular}[t]{cc||cc|cc}
\hline  \hline
   \multicolumn{2}{c||}{Decay Channels}   &  \multicolumn{2}{c|}{$B^0 \to a_1^+ b_1^-$}& \multicolumn{2}{c}{$B^0 \to b_1^+ a_1^-$} \\

 Parameter  & Definition & This work &   QCDF   & This work &   QCDF \\
\hline
  BR($10^{-6}$)        & $\Gamma/ \Gamma_{\rm total}$
  &$73.6^{+23.4+12.8+6.5}_{-17.0-12.1-6.0}$&
  $41.3^{+20.7+16.6}_{-18.2-3.4}$&
  $3.7^{+0.6+2.0+0.2}_{-0.5-1.6-0.2}$&
  $0.8^{+1.1+3.6}_{-0.4-0.1}$\\
 $f_L$      & $|{\cal A}_L|^2$
 &$0.94^{+0.00+0.00+0.00}_{-0.01-0.03-0.01}$&
  $0.90^{+0.02}_{-0.05}$&
  $0.96^{+0.01+0.01+0.00}_{-0.01-0.03-0.01}$&
  $0.98^{+0.00}_{-0.80}$\\
 $f_{||}$   & $|{\cal A}_{||}|^2$
 &$0.04^{+0.00+0.00+0.00}_{-0.00-0.02-0.00}$&
 &$0.02^{+0.01+0.02+0.00}_{-0.00-0.01-0.00}$
  \\
 $f_{\perp}$& $|{\cal A}_\perp|^2$
 &$0.03^{+0.00+0.00+0.00}_{-0.00-0.02-0.00}$&
 &$0.02^{+0.01+0.02+0.00}_{-0.00-0.01-0.00}$
 \\
 \hline \hline
\end{tabular}}
\end{center}
\end{table}
%% ==========================================================================================================

\noindent{\bf Branching ratios}
\bigskip

We first analyze the branching ratios of the $B \to a_1 b_1$ decays according to the
numerical results obtained in the pQCD approach. And furthermore, since these considered
modes have been studied in another popular QCDF approach, we also quote the related
predictions to make an essential comparison and discussion, which could be helpful
to further discriminate these two rather different tools through the future precise measurements.

As presented in Tables~\ref{tab:Br-Pf-p0-00}-\ref{tab:Br-Pf-pm-mp}, the pQCD
predictions for the branching ratios of the classified five modes
\footnote{It should be stressed that the final states in the former
$B^+ \to a_1^+ b_1^0$,
$B^+ \to b_1^+ a_1^0$, and $B^0 \to a_1^0 b_1^0$ modes are the {\it CP} eigenstates,
while those in the latter $B^0 \to a_1^+ b_1^-$ and
$B^0 \to b_1^+ a_1^-$ ones are not, which therefore result in the branching
ratios with and without the {\it CP}-averaged final states as presented
in Tables~\ref{tab:Br-Pf-p0-00}
and~\ref{tab:Br-Pf-pm-mp}, respectively.} are from $10^{-6}$
to $10^{-5}$, explicitly,
\beq
 \left.\begin{array}{lll}
 BR(B^+ \to a_1^+ b_1^0) &=& \hspace{0.16cm}9.0^{+5.5}_{-4.0} \times 10^{-6}\;,\\
 BR(B^+ \to b_1^+ a_1^0) &=& \hspace{0.16cm}4.2^{+2.1}_{-1.5} \times 10^{-6}\;, \\
 BR(B^0 \to a_1^0 b_1^0) &=& \hspace{0.16cm}3.3^{+1.9}_{-1.6} \times 10^{-6}\;; \\
 BR(B^0 \to a_1^+ b_1^-) &=& 73.6^{+27.5}_{-21.7} \times 10^{-6}\;,\\
 BR(B^0 \to b_1^+ a_1^-) &=& \hspace{0.16cm}3.7^{+2.1}_{-1.7} \times 10^{-6}\;; \\ \end{array} \right\}
\hspace{0.35cm}{\rm  (In \hspace{0.35cm}  pQCD) }
\label{eq:BR-a1b1-pQCD}
\eeq
%% ==========================================================================================================
\begin{table}[b]
\caption{Same as Table~\ref{tab:Br-Pf-p0-00} but
for the $B^0/\bar{B}^0\to a_1^+ b_1^-, B^0/\bar{B}^0\to b_1^+ a_1^-,
B^0 \to a_1^+ b_1^-+b_1^+ a_1^-$ decays.} \label{tab:Br-Pf-pmmp}
 \begin{center}\vspace{-0.5cm}{\footnotesize
\begin{tabular}[t]{cc||c|c|c}
\hline  \hline
   \multicolumn{2}{c||}{Decay Channels}   &  \multicolumn{1}{c|}{$B^0/\bar{B}^0\to a_1^+ b_1^-$}& \multicolumn{1}{c|}{$B^0/\bar{B}^0\to b_1^+ a_1^-$}& \multicolumn{1}{c}{$B^0 \to a_1^+ b_1^-+b_1^+ a_1^-$} \\
 Parameter  & Definition & This work &   This work   & This work \\
\hline
  BR($10^{-6}$)        & $\Gamma/ \Gamma_{\rm total}$
  &$91.1^{+29.1+20.7+9.1}_{-21.2-18.9-8.6}$&
  $44.2^{+18.0+8.1+3.6}_{-12.5-7.6-3.2}$&
  $85.8^{+24.3+19.3+6.3}_{-17.8-17.5-5.8}$\\
  \hline
 $f_L$      & $|{\cal A}_L|^2$
 &$0.91^{+0.01+0.05+0.01}_{-0.00-0.02-0.00}$&
  $0.81^{+0.02+0.07+0.01}_{-0.02-0.06-0.00}$&
  $0.91^{+0.00+0.03+0.00}_{-0.00-0.01-0.00}$\\
  $f_{||}$   & $|{\cal A}_{||}|^2$
 &$0.05^{+0.00+0.02+0.00}_{-0.00-0.02-0.00}$&
  $0.11^{+0.01+0.03+0.00}_{-0.01-0.05-0.00}$&
  $0.05^{+0.00+0.01+0.00}_{-0.00-0.01-0.00}$\\
 $f_{\perp}$& $|{\cal A}_\perp|^2$
 &$0.04^{+0.00+0.00+0.00}_{-0.01-0.03-0.01}$&
  $0.08^{+0.01+0.02+0.00}_{-0.01-0.02-0.00}$&
  $0.04^{+0.00+0.00+0.00}_{-0.00-0.01-0.00}$\\
 \hline \hline
\end{tabular}}
\end{center}
\end{table}
%%%========================================================================================================
%%%========================================================================================================
which are generally consistent with those estimated in the QCDF approach, namely,
\beq
 \left.\begin{array}{lll}
 BR(B^+ \to a_1^+ b_1^0) &=& 37.8^{+26.5}_{-16.2} \times 10^{-6}\;,\\
 BR(B^+ \to b_1^+ a_1^0) &=& \hspace{0.16cm}1.0^{+6.4}_{-0.5} \times 10^{-6}\;,\\
 BR(B^0 \to a_1^0 b_1^0) &=& \hspace{0.16cm}3.8^{+6.7}_{-2.4} \times 10^{-6}\;.\\
 BR(B^0 \to a_1^+ b_1^-) &=& 41.3^{+26.5}_{-18.5} \times 10^{-6}\;,\\
 BR(B^0 \to b_1^+ a_1^-) &=& \hspace{0.16cm}0.8^{+3.8}_{-0.4} \times 10^{-6}\;.\\ \end{array} \right\}
\hspace{0.35cm}{\rm  (In \hspace{0.35cm}  QCDF) }  \label{eq:BR-a1b1-QCDF}
\eeq
within still large theoretical errors. Notice that various errors here have been added in quadrature. All these predictions of the $B \to a_1 b_1$ decay rates with both QCDF and pQCD
approaches are expected to be accessed by the current LHCb and the forthcoming Belle-II
experiments.

As discussed in Refs.~\cite{Yang:2005gk,Yang:2007zt,Cheng:2007mx} with QCD sum rule method, relative to the vector $\rho$ meson, the two axial-vector $a_1$ and $b_1$ states
exhibit rather different hadron dynamics, namely, the former(latter) is similar(contrary) to $\rho$ with (anti)symmetric leading-twist distribution amplitude dominated by the
longitudinal(transverse) polarization.
Therefore, the involved QCD dynamics in the $B \to a_1 b_1$ decays should be
different from that in the $B \to a_1 a_1$ and $B \to b_1 b_1$ processes, while
similar to that in the $B \to b_1 \rho$ modes.
The $B \to a_1 a_1, b_1 b_1$ and $b_1 \rho$ channels have been investigated in the QCDF~\cite{Cheng:2008gxa}
and pQCD~\cite{Zhang:2012ew,Liu:2012jb} approaches.

Some remarks on the branching ratios of the $B \to a_1 b_1$ decays
are in order as follows:
\begin{itemize}
\item[]{(a)}
For the $B^+ \to a_1^+ b_1^0$ and $B^+ \to b_1^+ a_1^0$ decays, the branching
ratios predicted in the pQCD approach show different phenomena to those in the
QCDF approach, though the similar pattern of $Br(B^+ \to a_1^+ b_1^0) >
Br(B^+ \to b_1^+ a_1^0)$ has been gotten in terms of the central values. One
can clearly see from Eqs.~(\ref{eq:BR-a1b1-pQCD}) and~(\ref{eq:BR-a1b1-QCDF})
that $Br(B^+ \to a_1^+ b_1^0)_{\rm pQCD} \sim Br(B^+ \to b_1^+ a_1^0)_{\rm pQCD}$ while $Br(B^+ \to a_1^+ b_1^0)_{\rm QCDF} > Br(B^+ \to b_1^+ a_1^0)_{\rm QCDF}$
within errors.
The underlying reason is that the weak annihilation contributions paly an important
role in these two decays, which can be seen explicitly from the pQCD results of the
decay amplitudes given in the Table~\ref{tab:DA-p0-00} with different topologies.

\hspace{0.22cm}
Different from the $B^+ \to \rho^+ \rho^0, a_1^+ a_1^0,$ and $b_1^+ b_1^0$
decays, the large annihilation contributions appear in the $B^+ \to a_1^+ b_1^0$
and $b_1^+ a_1^0$ ones. Based on the assumption of the isospin symmetry,
the final states such as $\rho^+ \rho^0$, $a_1^+ a_1^0$, and $b_1^+ b_1^0$ are
the identical bosons, which, because of Bose-Einstein statistics, consequently lead to the exact cancellation between the
amplitudes induced by the $u\bar u$ and $d\bar d$ components of the neutral
state in the annihilation diagrams. However, the
$a_1$ and $b_1$ states are not the identical particles with different
quantum numbers.
The rather different QCD behaviors
between the $a_1$ and $b_1$ mesons further
result in the largely nonzero
annihilation decay amplitudes associated with the $a_1^+ b_1^0$ and
$b_1^+ a_1^0$
final states, respectively.

\hspace{0.22cm}
These two $B^+ \to a_1^+ b_1^0$ and $B^+ \to b_1^+ a_1^0$ decays with large
decay rates[${\cal O}(10^{-6})$] are expected to be tested by the LHCb and Belle-II experiments, which could, on one hand, confirm the reliability of the
perturbative calculations in the framework of pQCD or QCDF; on the other hand,
provide more evidences to distinguish the validity of the treatments
in calculating
the annihilation diagrams between the pQCD approach and SCET,
even to further understand the annihilation decay mechanism in the $B$ meson decays.

\item[]{(b)}
Analogous to $B^0 \to \rho^0 \rho^0, a_1^0 a_1^0$, and $b_1^0 b_1^0$ decays,
the $B^0 \to a_1^0 b_1^0$ channel is also dominated by the color-suppressed
tree amplitude. But, different from the small $Br(B^0 \to \rho^0 \rho^0)
\sim 0.3 \times 10^{-6}$ at leading order in the pQCD approach~\cite{Li:2006cva}, the
$B^0 \to a_1^0 b_1^0$ decay rate is about ten times
larger, which is slightly larger than the $B^0 \to a_1^0 a_1^0$ one
while almost one order less than the $B^0 \to b_1^0 b_1^0$ one in the pQCD approach~\cite{Liu:2012jb}. It is
interesting to note that this
phenomenon, i.e., $Br(B^0 \to a_1^0 a_1^0) < Br(B^0 \to a_1^0 b_1^0) <
Br(B^0 \to b_1^0 b_1^0)$, is attributed to the rather different QCD
behaviors between the $a_1$ and
$b_1$ mesons. Because of the extremely
small Wilson coefficient $a_2$ or
vanished decay constant $f_{b_1^0}$, then the $B^0 \to a_1^0 b_1^0$
decay amplitude will be determined by the nonfactorizable spectator
and annihilation diagrams. But, due to the great cancelation of the
annihilation
contributions, as can be seen in Table~\ref{tab:DA-p0-00}, the nonfactorizable
spectator amplitudes dominate the $B^0 \to a_1^0 b_1^0$ process. The underlying
reason is that the destructive(constructive) interferences between the
diagrams (c) and (d) in Fig.~\ref{fig:fig1} exhibit for the $a_1(b_1)$ emission
associated with the (anti)symmetric distribution amplitudes. Moreover,
the $B^0 \to a_1^0 a_1^0, a_1^0 b_1^0,$ and $b_1^0 b_1^0$ decay rates have also
been studied in the QCDF approach, which presented a different pattern, i.e.,
$Br(B^0 \to a_1^0 a_1^0) \lesssim Br(B^0 \to a_1^0 b_1^0) \sim
Br(B^0 \to b_1^0 b_1^0)$~\cite{Cheng:2008gxa}. These two different patterns among
the branching ratios of the $B^0 \to a_1^0 a_1^0, a_1^0 b_1^0,$ and $b_1^0 b_1^0$ decays in the pQCD and QCDF approaches would be tested by the near future experiments
due to their sizable values.

\item[]{(c)}
It is of great interest to note that the $B^0 \to a_1^+ b_1^-$ and
$B^0 \to b_1^+ a_1^-$ decays are dominated by the factorizable emission
contributions and nonfactorizable spectator amplitudes correspondingly.
Furthermore, for the former decay,
with the decay constant $f_{a_1} = 0.238$~GeV, a bit larger than that of the
$\rho$ meson, meanwhile, with the form factor $V_0^{B \to b_1} > V_0^{B \to
a_1}$, then the pattern $Br(B^0 \to a_1^+ b_1^-) > Br(B^0 \to a_1^+ a_1^-) >
Br(B^0 \to \rho^+ \rho^-)$ would be observed naturally.
But, for the latter mode, i.e., $B^0 \to b_1^+ a_1^-$, with $b_1^+$ emission,
because of the extremely suppressed decay constant $f_{b_1}\sim 0.0028$~GeV,
the factorizable emission diagrams give nearly zero contributions, which means
that the related decay amplitude might be induced by the nonfactorizable
spectator and weak annihilation diagrams if it could be detected in
the future. In fact, it is hopeful to
be measured at LHCb and/or Belle-II experiments in the near future in
light of its
large decay rate about $10^{-6}$ in the pQCD approach. Indeed, because
of the antisymmetric property of the $b_1$ meson twist-2 distribution
amplitude, then the constructive interferences between the diagrams
Fig.~\ref{fig:fig1}(c) and~\ref{fig:fig1}(d) 
lead to a dominant contribution to the 
$B^0 \to b_1^+ a_1^-$ mode, which can be seen from the values of the decay
amplitudes
shown in the Table~\ref{tab:DA-pm-mp-pmmp}.
As aforementioned, the nonfactorizable spectator and annihilation amplitudes
in the QCDF approach have to be fitted by the precision measurements due to
the endpoint singularities occurring in the collinear factorization theorem.
Therefore, this channel could act as one of the important roles to identify
the naive factorization breaking effects and distinguish the different factorization
approaches simultaneously.

\item[]{(d)} It should be stressed that the branching ratios shown in
Table~\ref{tab:Br-Pf-pm-mp} are not the {\it CP}-averaged ones. Actually,
the analyses of the $B^0 \to a_1^\pm b_1^\mp$  modes
are complicated because the involved final states are not the CP eigenstates.
Both $B^0$ and $\bar{B}^0$ mesons can decay into the same final states
simultaneously, i.e.,
$B^0/\bar{B}^0 \to a_1^+ b_1^-$ and $B^0 /\bar{B}^0 \to b_1^+ a_1^-$.
Due to $B^0-\bar{B}^0 $ mixing, it is very difficult to distinguish
$B^0$ from $\bar{B}^0 $. However, it is easy to identify the charged
final states in the considered decays.
We therefore sum up $B^0/\bar{B}^0\to a_1^+ b_1^- $ as one channel and
$B^0/\bar{B}^0\to b_1^+ a_1^- $ as another.
Meanwhile, following the convention adopted by the experimental
measurements~\cite{Olive:2016xmw},
we also define the {\it CP}-averaged channel as
$B^0 \to a_1^+ b_1^- + b_1^+ a_1^-$.
The numerical results for the branching ratios of these newly defined
channels are collected in the Table~\ref{tab:Br-Pf-pmmp}, specifically,
\beq
 BR(B^0/\bar{B}^0\to a_1^+ b_1^-) &=& 91.1^{+36.9}_{-29.7} \times 10^{-6}\;,\\
 BR(B^0/\bar{B}^0\to b_1^+ a_1^-) &=& 44.2^{+20.1}_{-15.0} \times 10^{-6}\;,\\
 BR(B^0 \to a_1^+ b_1^-+b_1^+ a_1^-) &=& 85.8^{+31.7}_{-25.6} \times 10^{-6}\;;
\label{eq:12}
 \eeq
Although the above-mentioned three channels are not
discussed in the QCDF
approach, the values predicted in the pQCD approach are such large that can be
easily accessed at the current LHCb and forthcoming Belle-II experiments. The
near future confirmations would help us to further explore the {\it CP}
violation, the CKM unitary angle $\alpha$, and so on in these interesting channels.

\item[]{(e)}
From the results presented in the
Tables~\ref{tab:Br-Pf-p0-00}-\ref{tab:Br-Pf-pmmp}, one can find that the
predicted branching ratios suffer from large theoretical uncertainties from the
not well-constrained meson wave functions in the considered decay modes.
To date, 
most of the $B \to AA$ decays are not measured yet,
except for the $B^0 \to a_1^+ a_1^-$ one observed by the {\it BABAR} Collaboration~\cite{Aubert:2009zr}. Therefore, we will define
some ratios among the branching ratios predicted in the pQCD approach by
adopting the $B^0 \to a_1^+ a_1^-$ decay rate
as the normalized one. Therefore, the related ratios are provided for
experimental
detection in the (near) future as follows:
\beq
R_{1} &\equiv&
\frac{BR(B^0 \to a_1^+ b_1^- + b_1^+ a_1^-)}{BR(B^0 \to a_1^+ a_1^-)}
\approx 1.57^{+0.23+0.60+0.02}_{-0.08-0.32-0.05}\;;
\eeq
%%%
\beq
R_{2} &\equiv&
\frac{BR(B^+ \to a_1^+ b_1^0)}{BR(B^0 \to a_1^+ a_1^-)}
\approx 0.17^{+0.03+0.00+0.01}_{-0.03-0.00-0.00}\;; \qquad
R_{3} \equiv
\frac{BR(B^+ \to b_1^+ a_1^0)}{BR(B^0 \to a_1^+ a_1^-)}
\approx 0.08^{+0.02+0.01+0.00}_{-0.02-0.01-0.00}\;;
\eeq
%%%
\beq
R_{4} &\equiv&
\frac{BR(B^0 \to a_1^0 b_1^0)}{BR(B^0 \to a_1^+ a_1^-)}
\approx 0.06^{+0.01+0.00+0.00}_{-0.01-0.00-0.00}\;;
\eeq
Moreover, we also define several ratios among the branching ratios themselves of the
$B \to a_1 b_1$ decays in this work as follows:
\beq
R_{5} &\equiv&
\frac{BR(B^0 \to b_1^+ a_1^-)}{BR(B^0 \to a_1^+ b_1^-)}
\approx 0.05^{+0.01+0.02+0.00}_{-0.01-0.02-0.00}\;; \qquad
R_{6} \equiv
\frac{BR(B^+ \to b_1^+ a_1^0)}{BR(B^+ \to a_1^+ b_1^0)}
\approx 0.46^{+0.04+0.07+0.02}_{-0.02-0.02-0.00}\;;
\eeq
%%%
\beq
R_{7} &\equiv&
\frac{BR(B^0 \to a_1^0 b_1^0)}{BR(B^+ \to a_1^+ b_1^0)}
\approx 0.37^{+0.01+0.00+0.00}_{-0.01-0.03-0.01}\;; \qquad
R_{8} \equiv
\frac{BR(B^0 \to a_1^0 b_1^0)}{BR(B^+ \to b_1^+ a_1^0)}
\approx 0.79^{+0.06+0.03+0.00}_{-0.07-0.15-0.03}\;;
\eeq
%%%
\beq
R_{9} &\equiv&
\frac{BR(B^+ \to a_1^+ b_1^0)}{BR(B^0 \to a_1^+ b_1^- + b_1^+ a_1^-)}
\approx 0.11^{+0.01+0.03+0.00}_{-0.00-0.02-0.01}\;;
\eeq
%%%
\beq
R_{10} &\equiv&
\frac{BR(B^+ \to b_1^+ a_1^0)}{BR(B^0 \to a_1^+ b_1^- + b_1^+ a_1^-)}
\approx 0.05^{+0.01+0.01+0.00}_{-0.01-0.01-0.00}\;;
\eeq
%%%
\beq
R_{11} &\equiv&
\frac{BR(B^0 \to a_1^0 b_1^0)}{BR(B^0 \to a_1^+ b_1^- + b_1^+ a_1^-)}
\approx 0.04^{+0.00+0.01+0.00}_{-0.00-0.01-0.00}\;;
\eeq
In the above ratios, the large uncertainties induced by the nonperturbative
inputs could be canceled to a great extent, which
are expected to
be measured in the future.
\end{itemize}
%%%%%%%%%%%%%%%%%%%%%%%%%%%%%%%%%%%%%%%%%%%%%%%%%%%%%%%%%%%%%%%%%%%%%%%%%
%%***********************************************************************
\begin{table}[!!h]
\caption{ The decay amplitudes(in unit of $10^{-3}\; \rm{GeV}^3$) of the
$B^+ \to a_1^+ b_1^0$, $B^+ \to b_1^+ a_1^0$, and $B^0 \to a_1^0 b_1^0$ channels
with three polarizations, where only the central values are quoted for clarification.}
\label{tab:DA-p0-00}
 \begin{center}\vspace{-0.3cm}{ \footnotesize
\begin{tabular}[t]{c||c|c|c|c|c|c|c|c}
\hline  \hline
  Channel   &  \multicolumn{8}{c}{$B^+ \to a_1^+ b_1^0$}\\
   \hline
 Decay Amplitudes & ${\cal A}^T_{fs}$ & ${\cal A}^P_{fs}$ &  ${\cal A}^T_{nfs}$ &  ${\cal A}^P_{nfs}$
                  & ${\cal A}^T_{nfa}$ &  ${\cal A}^P_{nfa}$ & ${\cal A}^T_{fa}$  & ${\cal A}^P_{fa}$\\
\hline \hline
 $L$      &$ 0.52 +{\it i} 1.50$ &$-0.13 + {\it i} 0.05$
          &$ 2.29 -{\it i} 0.97$ &$0.05 + {\it i} 0.12$
          &$0.37 -{\it i} 0.47$ &  $0.04 +{\it i} 0.02$
         &$0.01 -{\it i} 0.01$ &  $-0.51 +{\it i} 0.31$
 \\
 $N$      &$ 0.31 +{\it i} 0.89$ &  $-0.09 +{\it i} 0.04$
          &$-0.59 -{\it i} 0.02$ &  $ -0.01 - {\it i} 0.05$
          &$0.02 -{\it i} 0.02$ &  $\sim 0.00$
         &$\sim 0.00$ &  $0.08 +{\it i} 0.27$
 \\
 $T$      &$ -0.16 -{\it i} 0.47$ &  $0.07 -{\it i} 0.03$
          &$0.23 -{\it i} 0.31$ &  $ 0.02 +{\it i} 0.06$
          &$\sim 0.00$ &  $\sim 0.00$
         &$-0.25 -{\it i} 2.69$ &  $0.46 +{\it i} 0.49$
 \\
 \hline \hline
  Channel  & \multicolumn{8}{c}{$B^+ \to b_1^+ a_1^0$} \\
   \hline
 Decay Amplitudes & ${\cal A}^T_{fs}$ & ${\cal A}^P_{fs}$ &  ${\cal A}^T_{nfs}$ &  ${\cal A}^P_{nfs}$
                  & ${\cal A}^T_{nfa}$ &  ${\cal A}^P_{nfa}$ & ${\cal A}^T_{fa}$  & ${\cal A}^P_{fa}$\\
\hline \hline
 $L$      &$-0.03 -{\it i} 0.09$ &  $ -0.05 +{\it i} 0.02$
          &$ -0.62 +{\it i} 0.49$ &  $ -0.03 - {\it i} 0.06$
          &$-0.36 +{\it i} 0.47$ &  $ -0.04 -{\it i} 0.02$
          &$ -0.04 +{\it i} 0.02$ &  $ 0.50 - {\it i} 0.32$
 \\
 $N$      &$ -0.05 -{\it i} 0.14$ &  $-0.03 + {\it i} 0.01$
          &$0.66 -{\it i} 0.33$ &  $ 0.01 +{\it i} 0.05$
          &$ -0.02 +{\it i} 0.02$ &  $\sim 0.00$
          &$\sim 0.00$ &  $ -0.09 -{\it i} 0.28$
 \\
 $T$      &$ 0.08 +{\it i} 0.23$ &  $0.01 - {\it i} 0.01$
          &$-1.63 +{\it i} 0.16$ &  $-0.06 - {\it i} 0.07$
          &$ \sim 0.00$ &  $\sim 0.00$
          &$0.25 +{\it i} 2.69$ &  $-0.46 - {\it i} 0.49$
 \\
 \hline \hline
  Channel   &  \multicolumn{8}{c}{$B^0 \to a_1^0\, b_1^0$}\\
   \hline
 Decay Amplitudes & ${\cal A}^T_{fs}$ & ${\cal A}^P_{fs}$ &  ${\cal A}^T_{nfs}$ &  ${\cal A}^P_{nfs}$
                  & ${\cal A}^T_{nfa}$ &  ${\cal A}^P_{nfa}$ & ${\cal A}^T_{fa}$  & ${\cal A}^P_{fa}$\\
\hline \hline
 $L$      &$ 0.02 +{\it i} 0.06$ &  $-0.06 + {\it i} 0.02$
          &$-2.13 +{\it i} 0.50$ &  $ -0.02 - {\it i} 0.12$
          &$ 0.25 -{\it i} 0.33$ &  $0.04 - {\it i} 0.27$
          &$ -0.11 -{\it i} 0.07$ &  $ -0.07 + {\it i} 0.05$
 \\
 $N$      &$ 0.03 +{\it i} 0.09$ &  $-0.04 + {\it i} 0.02$
          &$ -0.13 +{\it i} 0.45$ &  $-0.03 - {\it i} 0.01$
          &$ \sim0.00  $ &  $ \sim 0.00$
          &$ 0.06 +{\it i} 0.18$ &  $0.15 - {\it i} 0.05$
 \\
 $T$      &$ -0.06 -{\it i} 0.17$ &  $0.05 - {\it i} 0.02$
          &$ 1.95 +{\it i} 0.17$ &  $-0.00 + {\it i} 0.13$
          &$ -0.10 +{\it i} 0.12 $ &  $  0.01+{\it i} 0.01$
          &$ \sim0.00 $ &  $\sim0.00 $
 \\
\hline  \hline
\end{tabular}}
\end{center}
\end{table}
%%%%%%%%%%%%%%%%%%%%%%%%%%%%%%%%%%%%%%%%%%%%%%%%%%%%%%%%%%%%%%%%%%%%%%%%%

\bigskip
\noindent{\bf Polarization fractions}
\bigskip

We now turn to the analyses of the polarization fractions.
Usually, the observables such as polarization fractions are
presented by
employing the transversity amplitudes. Then, based on the Eqs.~(\ref{eq:ase})
and (\ref{eq:norm}), the longitudinal polarization fraction can be defined as
\beq
f_L &\equiv& \frac{|{\cal A}_L|^2}{|{\cal A}_L|^2+|{\cal A}_\parallel|^2+|{\cal A}_\perp|^2}
= |{\cal A}_L|^2\;;
\label{eq:fl}
\eeq
The other two polarization fractions $f_{\parallel}$ and $f_{\perp}$ can be easily
obtained with similar definition to that shown in Eq.~(\ref{eq:fl}). One often use
another convention $f_{T}$, relative to $f_L$, to denote the transverse polarization fraction as,
\beq
f_{T} &\equiv& f_{\parallel} + f_\perp = 1- f_L \;;
\eeq

The polarization fractions predicted in both of the pQCD and QCDF approaches have been
collected in the Tables~\ref{tab:Br-Pf-p0-00}-\ref{tab:Br-Pf-pm-mp}. The longitudinal
and transverse polarization fractions can be read
as follows:
\beq
 \left.\begin{array}{lll}
 f_L(B^+ \to a_1^+ b_1^0) &=& 0.62^{+0.01}_{-0.05} \;, \qquad
 f_T(B^+ \to a_1^+ b_1^0) = 0.38^{+0.08}_{-0.04}\;; \\
 f_L(B^+ \to b_1^+ a_1^0) &=& 0.28^{+0.02}_{-0.04} \;, \qquad
 f_T(B^+ \to b_1^+ a_1^0) = 0.72^{+0.06}_{-0.09} \;;\\
 f_L(B^0 \to a_1^0 b_1^0) &=& 0.63^{+0.06}_{-0.10} \;, \qquad
 f_T(B^0 \to a_1^0 b_1^0) = 0.37^{+0.10}_{-0.07}\;; \\
 f_L(B^0 \to a_1^+ b_1^-) &=& 0.94^{+0.00}_{-0.03} \;, \qquad
 f_T(B^0 \to a_1^+ b_1^-) = 0.07^{+0.00}_{-0.03}\;;\\
 f_L(B^0 \to b_1^+ a_1^-) &=& 0.96^{+0.01}_{-0.03} \;, \qquad
 f_T(B^0 \to b_1^+ a_1^-) = 0.04^{+0.03}_{-0.01}\;; \\ \end{array} \right\}
\hspace{0.35cm}{\rm  (In \hspace{0.35cm}  pQCD) }
\label{eq:fp-a1b1-pQCD}
\eeq
\beq
 \left.\begin{array}{lll}
 f_L(B^+ \to a_1^+ b_1^0) &=& 0.92^{+0.02}_{-0.24}\;; \\
 f_L(B^+ \to b_1^+ a_1^0) &=& 0.73^{+0.12}_{-0.82}\;;\\
 f_L(B^0 \to a_1^0 b_1^0) &=& 0.98^{+0.01}_{-0.31}\;; \\
 f_L(B^0 \to a_1^+ b_1^-) &=& 0.90^{+0.02}_{-0.05}\;;\\
 f_L(B^0 \to b_1^+ a_1^-) &=& 0.98^{+0.00}_{-0.80}\;; \\ \end{array} \right\}
\hspace{0.35cm}{\rm  (In \hspace{0.35cm}  QCDF) }
\label{eq:fp-a1b1-QCDF}
\eeq
and
\beq
 \left.\begin{array}{lll}
 f_L(B^0/\bar{B}^0 \to a_1^+ b_1^-) &=& 0.91^{+0.05}_{-0.02} \;, \qquad
 f_T(B^0/\bar{B}^0 \to a_1^+ b_1^-) = 0.09^{+0.02}_{-0.04}\;; \\
 f_L(B^0/\bar{B}^0 \to b_1^+ a_1^-) &=& 0.81^{+0.07}_{-0.06} \;, \qquad
 f_T(B^0/\bar{B}^0 \to b_1^+ a_1^-) = 0.19^{+0.04}_{-0.06}\;;\\
 f_L(B^0 \to a_1^+ b_1^- + b_1^+ a_1^-) &=& 0.91^{+0.03}_{-0.01} \;, \qquad
 f_T(B^0 \to a_1^+ b_1^- + b_1^+ a_1^-) = 0.09^{+0.01}_{-0.01}\;; \\ \end{array} \right\}
\hspace{0.35cm}{\rm  (In \hspace{0.35cm}  pQCD) }
\label{eq:fp-a1b1pm-pQCD}
\eeq
in which various errors
have been added in quadrature. These predictions
in both pQCD and QCDF approaches need tests
by the related experiments in the future.
In light of these numerical results, generally speaking, the considered
$B \to a_1 b_1$ decays are dominated by the longitudinal polarization
contributions in the pQCD approach, except for the $B^+ \to b_1^+ a_1^0$
mode with $f_L \sim (24\%-30\%)$. It is very interesting to note that the
longitudinal polarization fraction $f_L$ of the $B^+ \to b_1^+ a_1^0$ decay
was estimated in the QCDF approach with quite large uncertainties, which can
possibly lead to a domination of the transverse polarization amplitudes.

According to the decay amplitudes from every topology of the $B \to a_1 b_1$
decays
as shown in the Tables~\ref{tab:DA-p0-00}-\ref{tab:DA-pm-mp-pmmp}, the
clarifications on those polarization fractions in the pQCD approach are
in more detail as follows:

\begin{itemize}
\item[]{(a)}
For the $B^+ \to a_1^+ b_1^0$
and $B^+ \to b_1^+ a_1^0$ decays, different from the $B^+ \to a_1^+ a_1^0$
and $B^+ \to b_1^+ b_1^0$ ones,
the largely nonvanishing transverse amplitudes contribute significantly from
the factorizable annihilation topology. Meanwhile, at the longitudinal
polarization, due to the antisymmetric leading twist distribution amplitude
of the emitted $b_1$ meson, the nonfactorizable spectator diagrams as shown
in Fig.~\ref{fig:fig1}(c) and~\ref{fig:fig1}(d) can interfere with each
other constructively accompanied with a large and positive Wilson coefficient
$C_2$ for the $B^+ \to a_1^+ b_1^0$ mode while with a much smaller and negative
Wilson coefficient $C_1$ for the $B^+ \to b_1^+ a_1^0$ one.
Consequently, the further constructive interferences between the factorizable
emission and nonfactorizable spectator amplitudes result
in the slightly dominant
longitudinal contribution to the $B^+ \to a_1^+ b_1^0$ decay.

\item[]{(b)}
As we know, the $B^0 \to \rho^0 \rho^0$ mode has a small longitudinal
polarization fraction in the pQCD approach at leading order~\cite{Chen:2006jz,Li:2006cva}.
Phenomenologically, this is attributed to the significant cancellation at
the longitudinal polarization between the factorizable emission and
nonfactorizable spectator decay amplitudes, which result in the well-known
color-suppressed tree amplitude $C$, quite small in magnitude. Because the
behavior of $a_1$ meson is similar to that of the $\rho$ meson, so the
polarization fractions of $B^0 \to a_1^0 a_1^0$ decay~\cite{Liu:2012jb}
is also analogous
to those of the $B^0 \to \rho^0 \rho^0$ one. In other words,
the large transverse
decay amplitudes still exist. While, for the $B^0 \to a_1^0 b_1^0$ channel,
the aforementioned enhancement of the nonfactorizable spectator amplitudes
associated with the $b_1$ emission governs the longitudinal helicity amplitude
and finally results in the different polarization fractions to those of the
$B^0 \to \rho^0 \rho^0$ and $a_1^0 a_1^0$ decays. Therefore,
one can observe an interesting relation of the longitudinal polarization
fractions in the pQCD approach at leading order, that is, $f_L(B^0 \to a_1^0
a_1^0) < f_L(B^0 \to a_1^0 b_1^0) < f_L(B^0 \to b_1^0 b_1^0)$,
whose confirmation would provide more information to explore the
least understood quantity~\cite{Li:2009wba},
namely, the color-suppressed tree amplitude $C$,
in the $B$ physics.

\item[]{(c)}
As shown in the Table~\ref{tab:DA-pm-mp-pmmp}, both of the $B^+ \to a_1^+ b_1^-$
and $B^+ \to b_1^+ a_1^-$ decays are highly dominated by the longitudinal
polarization amplitudes but with different sources. The former decay has
a large color-allowed tree amplitude mainly arising from the factorizable
emission diagrams with Wilson coefficient $a_1$(not to be confused with
the abbreviation $a_1$ for the $a_1(1260)$ state). However, the latter
one has a bit smaller tree amplitude induced by the nonfactorizable spectator
diagrams
with Wilson coefficient $C_1$. Therefore, 
the $B^0 \to
a_1^+ b_1^- + b_1^+ a_1^-$ decay with {\it CP} eigenstate is
certainly
dominated by the longitudinal
polarization amplitude, which gives a large fraction around $90\%$.

\end{itemize}

\bigskip
\noindent{\bf Naive factorization breaking effects: nonfactorizable spectator
 and/or weak annihilation contributions}
\bigskip

Now, we will discuss the naive factorization breaking effects, that is, the
nonfactorizable spectator and/or weak annihilation diagrams contribute to the
above mentioned observables in the $B \to a_1 b_1$ decays.

It is well known that the naive factorization hypothesis
has been successfully applied
into various decay modes of heavy mesons and, particularly, the obtained
branching ratios for the color-allowed processes governed by the factorizable
contributions agree well with the data
generally. However, for the modes belonging to the color-suppressed category~\cite{Neubert:2001sj}
such as $B \to J/\psi K^{(*)}$(e.g., see~\cite{Cheng:1996xy,Cheng:1998kd,Chen:2005ht,Liu:2013nea}),
$B^0 \to \pi^0 \pi^0$(e.g., see~\cite{Chiang:2004nm,Charng:2004ed,Fleischer:2007mq,Li:2009wba,Liu:2015sra}), etc., the decay rates estimated in the naive factorization
are always too small to be compared with the measurements due to the nearly vanishing
Wilson coefficient $a_2 \sim 0$. Then the nonfactorizable
spectator even weak annihilation amplitudes should be included to clarify the
experimental measurements, although they
are usually considered as higher order(or power) corrections
contributing less in the naive factorization.

%%%%%%%%%%%%%%%%%%%%%%%%%%%%%%%%%%%%%%%%%%%%%%%%%%%%%%%%%%%%%%%%%%%%%%%%%
%%***********************************************************************
\begin{table}[htb]
\caption{ The decay amplitudes(in unit of $10^{-3}\; \rm{GeV}^3$) of the
$B^0 \to a_1^+ b_1^-$, $B^0 \to b_1^+ a_1^-$, and $B^0 \to  a_1^+ b_1^- + b_1^+ a_1^-$ channels with three polarizations, where only the central values are quoted for clarification.}
\label{tab:DA-pm-mp-pmmp}
 \begin{center}\vspace{-0.3cm}{ \footnotesize
\begin{tabular}[t]{c||c|c|c|c|c|c|c|c}
\hline  \hline
  Channel   &  \multicolumn{8}{c}{$B^0 \to a_1^+ b_1^-$}\\
   \hline
 Decay Amplitudes & ${\cal A}^T_{fs}$ & ${\cal A}^P_{fs}$ &  ${\cal A}^T_{nfs}$ &  ${\cal A}^P_{nfs}$
                  & ${\cal A}^T_{nfa}$ &  ${\cal A}^P_{nfa}$ & ${\cal A}^T_{fa}$  & ${\cal A}^P_{fa}$\\
\hline \hline
 $L$      &$ 3.84 +{\it i}11.03$ &  $-1.06 + {\it i} 0.44$
          &$-0.32 -{\it i} 0.13$ &  $ 0.01 - {\it i} 0.03$
          &$ 0.84 -{\it i} 1.16$ &  $-0.05- {\it i} 0.27$
          &$ -0.02 -{\it i} 0.04 $              &  $ -0.60 - {\it i} 0.21$
 \\
 $N$      &$ 0.53 +{\it i} 1.53$ &  $-0.15 + {\it i} 0.06$
          &$ -0.04 +{\it i} 0.30$ &  $-0.03 + {\it i} 0.00$
          &$ 0.02 +{\it i} 0.00$ &  $-0.00 + {\it i} 0.01$
          &$ \sim 0.00$               &  $0.29 + {\it i} 0.21$
 \\
 $T$      &$ 1.00 +{\it i} 2.87$ &  $-0.29 + {\it i} 0.12$
          &$ -0.06 +{\it i} 0.65$ &  $-0.06 - {\it i} 0.01$
          &$  0.01 -{\it i} 0.00$            &  $ -0.01 -{\it i} 0.01$
          &$-0.01 -{\it i} 0.00$ &  $ 0.58 +{\it i} 0.42$
 \\
 \hline \hline
  Channel  & \multicolumn{8}{c}{$B^0 \to b_1^+ a_1^-$} \\
   \hline
 Decay Amplitudes & ${\cal A}^T_{fs}$ & ${\cal A}^P_{fs}$ &  ${\cal A}^T_{nfs}$ &  ${\cal A}^P_{nfs}$
                  & ${\cal A}^T_{nfa}$ &  ${\cal A}^P_{nfa}$ & ${\cal A}^T_{fa}$  & ${\cal A}^P_{fa}$\\
\hline \hline
 $L$      &$-0.01 -{\it i} 0.02$ &  $ \sim 0.00$
          &$-1.57 +{\it i} 0.43$ &  $-0.06 - {\it i} 0.10$
          &$ -0.31 +{\it i} 0.38$ &  $-0.03 - {\it i} 0.30$
          &$-0.02 -{\it i} 0.04$ &  $0.57 + {\it i} 0.23$
 \\
 $N$      &$ \sim 0.00$ &  $\sim 0.00$
          &$ -0.06 -{\it i} 0.05$ &  $ \sim 0.00$
          &$-0.02 -{\it i} 0.00$ &  $-0.01 +{\it i} 0.01$
          &$ \sim 0.00$               &  $ -0.30 - {\it i} 0.21$
 \\
 $T$      &$ -0.00 -{\it i} 0.01$ &  $\sim 0.00$
          &$ -0.10 -{\it i} 0.08$ &  $ \sim 0.00$
          &$ 0.10 -{\it i} 0.00$               &  $ -0.12 -{\it i} 0.22$
          &$ 0.01 +{\it i} 0.00$ &  $ -0.58 - {\it i} 0.42$
 \\
 \hline \hline
Channel  & \multicolumn{8}{c}{$B^0 \to a_1^+ b_1^- + b_1^+ a_1^-$} \\
   \hline
 Decay Amplitudes & ${\cal A}^T_{fs}$ & ${\cal A}^P_{fs}$ &  ${\cal A}^T_{nfs}$ &  ${\cal A}^P_{nfs}$
                  & ${\cal A}^T_{nfa}$ &  ${\cal A}^P_{nfa}$ & ${\cal A}^T_{fa}$  & ${\cal A}^P_{fa}$\\
\hline \hline
 $L$      &$3.83 +{\it i} 11.01$ &  $-1.06 + {\it i} 0.44$
          &$-1.89 +{\it i} 0.30$ &  $-0.05 - {\it i} 0.13$
          &$ 0.53 -{\it i} 0.78$ &  $-0.08 - {\it i} 0.57$
          &$-0.04 -{\it i} 0.08$ &  $-0.03 + {\it i} 0.02$
 \\
 $N$      &$ 0.53 +{\it i} 1.53$ &  $-0.15 + {\it i} 0.06$
          &$ -0.10 +{\it i} 0.25$ &  $-0.03 + {\it i} 0.00$
          &$ \sim 0.00$ &  $-0.01 +{\it i} 0.02$
          &$ \sim 0.00$               &  $ -0.01 - {\it i} 0.00$
 \\
 $T$      &$ 1.00 +{\it i} 2.86$ &   $-0.29 + {\it i} 0.12$
          &$ -0.16 +{\it i} 0.57$ &   $-0.06 - {\it i} 0.01$
          &$ 0.11 -{\it i} 0.00$               &  $ -0.13 -{\it i} 0.23$
          &$ \sim 0.00$ & $ \sim 0.00$
 \\
 \hline \hline
\end{tabular}}
\end{center}
\end{table}
%%%%%%%%%%%%%%%%%%%%%%%%%%%%%%%%%%%%%%%%%%%%%%%%%%%%%%%%%%%%%%%%%%%%%%%%%
%%***********************************************************************

In order to simply investigate the naive factorization breaking effects,
we here just explore the branching ratios and longitudinal polarization
fractions in the considered modes when the nonfactorizable spectator and/or
annihilation contributions are turned off. For the sake of simplicity,
only the central values of the related observables are quoted here for
clarifications.
\begin{enumerate}
\item
When we neglect the contributions from the weak annihilation diagrams,
the decay rates and polarization fractions will become
\beq
Br(B^+ \to a_1^+ b_1^0) &\approx& 5.3 \times 10^{-6}\;, \qquad
f_L(B^+ \to a_1^+ b_1^0)\approx 0.81\;;\\
Br(B^+ \to b_1^+ a_1^0) &\approx& 1.5 \times 10^{-6}\;, \qquad
f_L(B^+ \to b_1^+ a_1^0)\approx 0.21 \;;\\
Br(B^0 \to a_1^0 b_1^0) &\approx& 3.7 \times 10^{-6}\;, \qquad
f_L(B^0 \to a_1^0 b_1^0)\approx 0.71\;;\\
Br(B^0 \to a_1^+ b_1^-) &\approx& 75.5 \times 10^{-6}\;,\qquad
f_L(B^0 \to a_1^+ b_1^-)\approx 0.91\;;\\
Br(B^0 \to b_1^+ a_1^-) &\approx& 1.4 \times 10^{-6}\;, \qquad
f_L(B^0 \to b_1^+ a_1^-)\approx 0.99\;;
\eeq
One can observe that the weak annihilation amplitudes contribute constructively
to the decay rates of the $B^+ \to a_1^+ b_1^0$, $B^+ \to b_1^+ a_1^0$, and
$B^0 \to a_1^+ b_1^-$ modes around $41\%$, $64\%$, and $58\%$, respectively,
however, destructively to those of the $B^0 \to a_1^0 b_1^0$ and $B^0 \to
a_1^+ b_1^-$ ones about $12\%$ and $3\%$, respectively. Moreover, the weak
annihilation contributions, in particular, the large factorizable annihilation
amplitudes, decrease the longitudinal polarization fraction nearly $31\%$
of the $B^+ \to a_1^+ b_1^0$ decay while increase that about $25\%$ of the
$B^+ \to b_1^+ a_1^0$ one. And an enhancement to the transverse polarization
fraction of the $B^0 \to a_1^0 b_1^0$ channel around $12\%$ can be easily seen
because of a bit large nonfactorizable annihilation contributions. The
polarization fractions only vary with 0.03 for the $B^0 \to a_1^+ b_1^-$
and $B^0 \to b_1^+ a_1^-$ decays with neglecting the annihilation amplitudes
since these two modes
are governed by the factorizable emission and nonfactorizable spectator
diagrams correspondingly. Nevertheless, one can still observe the significant
naive factorization breaking effects in the $B^+ \to a_1^+ b_1^0$,
$B^+ \to b_1^+
a_1^0$, and $B^0 \to a_1^+ b_1^-$ decays induced by the annihilation diagrams,
though which usually are regarded as being negligible due to its power suppression.

\item
Without the nonfactorizable spectator and weak annihilation contributions, then the branching ratios and the polarization fractions will become
\beq
Br(B^+ \to a_1^+ b_1^0) &\approx& 2.3 \times 10^{-6}\;, \qquad
f_L(B^+ \to a_1^+ b_1^0)\approx 0.58\;;\\
Br(B^+ \to b_1^+ a_1^0) &\approx& 4.2 \times 10^{-8}\;, \qquad
f_L(B^+ \to b_1^+ a_1^0)\approx 0.16 \;;\\
Br(B^0 \to a_1^0 b_1^0) &\approx& 2.2 \times 10^{-8}\;, \qquad
f_L(B^0 \to a_1^0 b_1^0)\approx 0.16\;;\\
Br(B^0 \to a_1^+ b_1^-) &\approx& 76.4 \times 10^{-6}\;,\qquad
f_L(B^0 \to a_1^+ b_1^-)\approx 0.94\;;\\
Br(B^0 \to b_1^+ a_1^-) &\approx& 2.4 \times 10^{-10}\;, \qquad
f_L(B^0 \to b_1^+ a_1^-)\approx 0.80\;.
\eeq
Relative to the naive factorization, when the so-called factorization breaking
terms are removed, then the considered $B \to a_1 b_1$ decays show different
phenomena in light of the branching ratios: the numerical results of $Br(B^+
\to b_1^+ a_1^0)$, $Br(B^0 \to a_1^0 b_1^0)$, and $Br(B^0 \to b_1^+ a_1^-)$
change from $10^{-6}$ to $10^{-8}$, even $10^{-10}$, which indicate evidently
that these modes are governed
by the naive factorization breaking effects. Therefore, it is proposed that
these processes could be detected by the relevant experiments in the (near)
future to verify those phenomenologies induced by the naive factorization
breaking effects. Of course, the $B^0 \to a_1^+ b_1^-$ mode is also an
ideal candidate
with a much large decay rate to test the naive factorization due to its extreme
dominance of the factorizable emission diagrams.
\end{enumerate}

%%%%%%%%%%%%%%%%%%%%%%%%%%%%%%%%%%%%%%%%%%%%%%%%%%%%%%%%%%%%%%%%%%%%%%
%%*******************************************************************
Finally, frankly speaking, the theoretical predictions in both of the pQCD
and QCDF approaches still have large uncertainties arising from various
sources. In terms of the pQCD approach, the theoretical errors mainly
come from the not well-constrained input parameters involved in the hadron
distribution amplitudes such as the shape parameter $\omega_B$ of heavy $B$
meson and the Gegenbauer moments $a_i^{\parallel,\perp}$ of light axial-vector
$a_1$ and $b_1$ states. Therefore, the great efforts from nonperturbative
QCD aspects such as QCD sum rule and/or Lattice QCD methods, as well as
from the experimental aspects, are eagerly desired to effectively
reduce the errors of these important inputs. Certainly, any progress
of the hadron dynamics would improve the precision of the predictions more
or less in the pQCD approach

%%%%%%%%%%%%%%%%%%%%%%%%%%%%%%%%%%%%%%%%%%%%%%%%%%%%%%%%%%%%%%%%%%%%%%
%%*******************************************************************

\bigskip
%\section{Summary}\label{sec:sum}

In summary, because of
the dramatically small or vanishing decay constant $f_{b_1}$ of the
light axial-vector $b_1$ state, the naive factorization would
provide an extremely small or nearly zero branching ratios, for
example, the $B^0 \to b_1^+ a_1^-$ mode. However, as indicated
from data, many processes may have large branching ratios since
the large naive factorization breaking effects such as nonfactorizable
spectator and/or annihilation contributions could exist.
Therefore, we should go beyond the naive factorization to explore
those possibly large factorization breaking effects.

We investigated the
branching ratios and polarization fractions of the charmless hadronic $B \to
a_1 b_1$ decays by employing the pQCD approach based
on the $k_T$ factorization theorem, with which we perturbatively calculated
the factorizable emission, nofactorizable spectator,
and weak annihilation diagrams.
The predicted branching ratios as large as
$10^{-5}-10^{-6}$ in the pQCD approach are
in general consistency with those estimated in the QCDF approach
within still large theoretical errors. Due to the antisymmetric
behavior of the $b_1$ meson leading twist distribution amplitude,
the nonfactorizable spectator contributions with $b_1$ emission
can change from destruction
into construction, which provide a large naive factorization breaking term
and further enhance the decay amplitudes significantly.
The predicted polarization
fractions in the pQCD approach are also consistent with those
given in the QCDF approach.

The detailed analyses show that the pQCD predictions of the
considered $B \to a_1 b_1$ decays could provide more evidences to test
the SM, explore the helicity structure with polarizations, constrain the
parameters from the hadron wave functions, and so forth. The large
$B^0 \to a_1^0 a_1^0, a_1^0 b_1^0,$ and $b_1^0 b_1^0$ decay rates
would provide an opportunity to make further constraints to the CKM unitary
angles and understandings of the decay mechanism of the color-suppressed
modes. Certainly, it is worth stressing that we firstly consider the short-distance
contributions at leading order in the evaluations of the hadronic matrix element of
the $B \to a_1 b_1$ decays.
The effects of final state interaction might play an important
role in these considered
processes as they should. However, it is beyond the scope
of the present work and will be studied elsewhere.

\bigskip
%\begin{acknowledgments}

This work is supported by the National Natural Science
Foundation of China under Grants No.~11765012, No.~11775117,
No.~11205072, and No.~11235005 and by the Research Fund of
Jiangsu Normal University under Grant No.~HB2016004.

%\end{acknowledgments}

%%%%%%%%%%%%%%%%%%%%%%%%%%%%%%%%%%%%%%%%%%%%%%%%%%%%%%%%%%%%%%%%%%%%%%%%%%%%%%%%%%
%                                 reference
%%%%%%%%%%%%%%%%%%%%%%%%%%%%%%%%%%%%%%%%%%%%%%%%%%%%%%%%%%%%%%%%%%%%%%%%%%%%%%%%%%

\end{document}